\documentclass[9pt,twocolumn,twoside]{osajnl}

\journal{ol_SB} 

\setboolean{shortarticle}{true}
  
\title{First-order perturbation theory for material changes in the surrounding of open optical resonators}

\author[1,*]{S.Both}
\author[1]{T.Weiss}

\affil[1]{4th Physics Institute and Research Center SCoPE, University of Stuttgart, Pfaffenwaldring 57, D-70550 Stuttgart, Germany}
\affil[*]{Corresponding author: s.both@pi4.uni-stuttgart.de}

\doi{}

\begin{abstract}
The single-mode approximation of the resonant state expansion has proven to give accurate first-order approximations of resonance shifts and linewidth changes when modifying the material properties inside open optical resonators. Here, we extend this first-order perturbation theory to modifications of the material properties in the surrounding medium. As a side product of our derivations, we retrieve the already known analytical normalization condition for resonant states. We apply our theory to two example systems: A metallic nanosphere and a one-dimensional photonic crystal slab. 
\end{abstract}

\setboolean{displaycopyright}{false}

\usepackage{amsmath, amssymb, bm}
\usepackage{txfonts} 
\usepackage{braket} 

\newcommand{\m}{ m } 
\newcommand{\p}{ \nu } 
\newcommand{\E}{ \mathbf{E} } 
\newcommand{\Hv}{ \mathbf{H} } 
\newcommand{\M}{ \hat{\varmathbb{M}} }
\newcommand{\Pop}{ \hat{\varmathbb{P}} }
\newcommand{\D}{ \hat{\varmathbb{D}} }
\newcommand{\F}{ \varmathbb{F} }
\newcommand{\A}{ \varmathbb{A} }
\newcommand{\B}{ \varmathbb{B} }
\newcommand{\J}{ \varmathbb{J} }
\newcommand{\basis}{ \varmathbb{O} } 
\newcommand{\rv}{ \mathbf{r} }
\newcommand{\epst}{ \hat{\bm{\varepsilon}} } 
\newcommand{\mut}{ \hat{\bm{\mu}} } 
\newcommand{\R}{ \text{R} } 
\newcommand{\N}{ \mathbf{N} } 
\newcommand{\ns}{ n_\text{S} } 
\newcommand{\es}{ \varepsilon_\text{S} } 
\newcommand{\mus}{ \mu_\text{S} } 
\newcommand{\fE}{ \mathbf{e} } 
\newcommand{\fH}{ \mathbf{h}  } 
\newcommand{\sV}{ {} } 
\newcommand{\spV}{ {} } 
\newcommand{\STl}{ S_{\spV} } 

\begin{document}

\maketitle

\begin{figure}[htb]
	\centering
	\includegraphics[width=\linewidth]{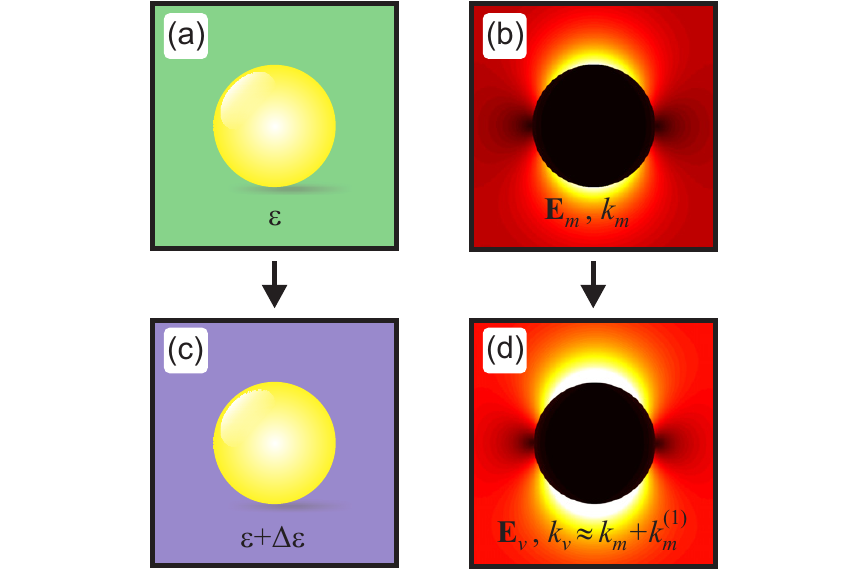}
	\caption{ Influence of the surrounding medium on the resonant states of an open optical system. Depicted is a metallic nanosphere, which also serves as our test system~(i). (a) In the unperturbed case, the sphere is surrounded by a medium with permittivity $\varepsilon$. (b) Exemplary resonant state of the unperturbed system, characterized by its electric field distribution $\E_\m$ and its vacuum wavenumber $k_\m$. (c) Perturbed system with permittivity $\varepsilon+\Delta\varepsilon$. (d) Resonant state of the perturbed system, characterized by a modified field distribution $\E_\p$ and a modified wavenumber $k_\p \approx k_\m+k_\m^{(1)}$.}
	\label{fig:sphere_drawing}
\end{figure}

Nanophotonic structures such as photonic crystals or plasmonic nanoparticles compromise optical resonances with strong electromagnetic near-fields. Consequently, even tiny changes in the the surrounding materials can have significant influence on the resonances frequencies. This is the key to various kinds of optical sensing applications~\cite{wilson2002, arnold2003, unger2009, liu2010, cetin2014, gallinet2013, meschFano2018}. Fig.~\ref{fig:sphere_drawing} displays exemplarily a metallic sphere, around which the surrounding permittivity is changed from $\varepsilon$ to $\varepsilon+\Delta\varepsilon$, thus shifting the resonance wavenumber from $k_\m$ to $k_\p$. 

The modeling of such systems often relies on extensive numerical simulations, which can be rather inefficient, since in many practical cases, the variations in the material properties are extremely small. In contrast, perturbative theories are particularly suited for these cases. They are based on the eigenmodes of the system, also known as resonant states (RS) or quasi-normal modes~\cite{lalanne2015simple, hughes2018reg, lalanne2018review, colom2018}, and have proven to be very efficient for all kinds of perturbations inside or in close proximity to nanophotonic resonators~\cite{DoostRSE3D2014, zhang2015univ, lalanne2015simple, mul2016sand, weiss2016perturbation, weiss2017oblique, upendar2018}. However, a general rigorous way to incorporate perturbations of the surrounding medium into the theory is missing so far. The main difficulty arises from the fact that nanophotonic systems exhibit RS that radiate to the far field, so that their field distributions grow with distance to the resonator~\cite{koenderink2010purcell, weiss2016perturbation, weiss2017oblique}. Hence, conventional perturbative formulations for bound states, e.g., known from quantum mechanics, cannot be applied. Several normalization schemes have been developed in recent years (for details, see Refs.~\cite{hughes2018reg, lalanne2018review, mul2018general} and references therein), but no theory exists so far for perturbations in the exterior. In this Letter, we derive such a theory for homogeneous and isotropic perturbations.

The frequency representation of Maxwell's equations [Gaussian units, time dependence $\exp(-i\omega t)$] can be written as~\cite{mul2018general}
\begin{equation}
	\M(k, \rv) \F(k, \rv)
	= 
	\J(k, \rv)
	\label{eq:maxwell},
\end{equation}
where the electric and magnetic fields, $\mathbf{E}$ and $\mathbf{H}$, as well as the electric current $\mathbf{j}$, are summarized in six-dimensional supervectors
\begin{equation}
	\F(k, \rv)
	= 
	\begin{pmatrix}
	\E(k, \rv)\\
	i\Hv(k, \rv)\\
	\end{pmatrix}
	\quad \text{and} \quad
	\J(k, \rv)
	= 
	\begin{pmatrix}
	-\frac{4\pi i}{c} \mathbf{j}(k, \rv)\\
	0\\
	\end{pmatrix},
	\label{eq:FandJdef}
\end{equation}
and $\M(k, \rv) = k \Pop(k, \rv) - \D(\rv)$, with
\begin{equation}
	\Pop(k, \rv)
	= 
	\begin{pmatrix}
		\epst(k, \rv) & 0\\
		0 & \mut(k, \rv)\\
	\end{pmatrix}
    \quad \text{and} \quad
	\D(\rv)
	= 
	\begin{pmatrix}
		0 & \nabla \times\\
		\nabla \times & 0\\
	\end{pmatrix}.
	\label{eq:PandDdef}
\end{equation}
For brevity of notation, we use wavenumbers $k=\omega/c$ instead of frequencies $\omega$. In the most general case, the operator $\Pop$ can also include bi-anisotropic materials~\cite{mul2018general}, which is however beyond the scope of this work. The RSs are defined as the solutions of \eqref{eq:maxwell} with outgoing boundary conditions in the absence of sources:
\begin{equation}
	\M(k_\m, \rv) \F_\m(\rv)
	= 
	0
	\label{eq:RS0definition},
\end{equation}
where $k_\m$ is the corresponding resonance wavenumber. Note that in open systems, $k_\m$ is complex valued, with $\mathrm{Re}(ck_\m)$ representing the resonance frequency and $-2\mathrm{Im}(ck_\m)$ specifying the linewidth. Now, let us introduce a perturbation $\lambda k \Delta\Pop(k, \rv)$ with
\begin{equation}
\Delta\Pop(k, \rv)
= 
\begin{pmatrix}
\Delta\epst(k, \rv) & 0\\
0 & \Delta\mut(k, \rv)\\
\end{pmatrix}
\label{eq:deltaPdef},
\end{equation}
where the perturbation parameter $\lambda$ allows to switch the perturbation on and off. The RSs of the perturbed system are then defined by
\begin{equation}
	\left[ \M(k_\p, \rv) + \lambda k_\p \Delta\Pop(k_\p, \rv) \right]  \F_\p(\rv)
	= 0
	\label{eq:RSdefinition},
\end{equation}
and are characterized by the modified wavenumber $k_\p$. In order to be as general as possible, we allow the quantities $\epst(k, \mathbf{r})$, $\Delta \epst(k, \mathbf{r})$, $\mut(k, \mathbf{r})$ and $\Delta \mut(k, \mathbf{r})$ to be  tensors that are dispersive and depend on $\rv$. We make only one restriction for the following derivations: We require that there is a homogeneous and isotropic surrounding, in which those quantities are represented by a spatially constant scalar value that can be written as $\epst(k, \mathbf{r}) = \mathbb{1}\varepsilon(k)$, $\Delta \epst(k, \mathbf{r}) = \mathbb{1}\Delta\varepsilon(k)$, $\mut(k, \mathbf{r}) = \mathbb{1}\mu(k)$, and $\Delta\mut(k, \mathbf{r}) = \mathbb{1}\Delta\mu(k)$, where $\mathbb{1}$ denotes a $3\times3$ unit matrix.

For later convenience, we introduce the following two bilinear maps~\cite{weiss2018scat}: For two six-dimensional supervectors 
\begin{equation}
	\A = \begin{pmatrix} \mathbf{A}_\text{E}\\ i\mathbf{A}_\text{H} \end{pmatrix}
	\quad\mathrm{and}\quad 
	\B = \begin{pmatrix} \mathbf{B}_\text{E}\\ i\mathbf{B}_\text{H} \end{pmatrix},
\end{equation}
we define a volume integral over a finite volume $V$ as
\begin{equation}
	\braket{\A|\B}_\sV
	\equiv 
	\int_V \mathrm{d}V
	\left( \mathbf{A}_\text{E} \cdot \mathbf{B}_\text{E} - \mathbf{A}_\text{H} \cdot \mathbf{B}_\text{H} \right)
	\label{eq:volumeIntegral}
\end{equation}
and a surface integral over the boundary $\partial V$ of $V$ as
\begin{equation}
	[\A|\B]_{\spV}
	\equiv 
	i\oint_{\partial V } \mathrm{d}\mathbf{S}
	\left( \mathbf{A}_\text{E} \times \mathbf{B}_\text{H} - \mathbf{B}_\text{E} \times \mathbf{A}_\text{H} \right).
	\label{eq:surfaceIntegral}
\end{equation}

In the following, we derive an expression that relates the unperturbed and the perturbed RSs. We introduce the superscript $\R$ for the reciprocal conjugate~\cite{weiss2018scat}, in order to label operators and fields that are evaluated at the same wavenumber $k$, but for reciprocal boundary conditions. For example, in planar periodic systems, this corresponds to an inversion of the in-plane momentum $\mathbf{k}_\parallel \rightarrow -\mathbf{k}_\parallel$~\cite{weiss2017oblique, weiss2018scat}. We multiply \eqref{eq:RSdefinition} with $\F^\R_\m$ and the reciprocal conjugate of \eqref{eq:RS0definition} with $\F_\p$ from the left, subtract both expressions, and use that $\M=\M^\R$ and $\Delta\Pop=\Delta\Pop^\R$. Then, we integrate the resulting equation over a finite volume $V$ enclosing the inhomogeneities of $\epst(k, \mathbf{r})$, $\Delta \epst(k, \mathbf{r})$, $\mut(k, \mathbf{r})$ and $\Delta \mut(k, \mathbf{r})$, and exploit the identities provided in Refs.~\cite{mul2018general, weiss2018scat}. This gives
\begin{equation}
	\begin{split}
	T(\lambda) 
	\equiv
	k_\p \braket{\F_\m^\R| \Pop(k_\p) |\F_\p}_\sV
	-
	k_\m \braket{\F_\p| \Pop(k_\m) |\F_\m^\R}_\sV
	\\
	+
	k_\p \braket{\F_\m^\R| \lambda \Delta\Pop(k_\p)|\F_\p}_\sV
	+
	[\F_\m^\R | \F_\p]_{\spV} = 0
	\label{eq:basicEq}.
	\end{split}
\end{equation}
As in standard perturbation theories, $T$, $\F_\p$, and $k_\p$ can be written as
\begin{gather}
		T(\lambda) = T(0) + \lambda \left.\frac{\mathrm{d}T}{\mathrm{d}\lambda}\right|_{\lambda=0} + \mathcal{O}(\lambda^2)  = 0, 		\label{eq:ansatzT} \\
		\F_\p = \F_\m + \lambda \F^{(1)}_\m + \mathcal{O}(\lambda^2), 
		\text{ and }
		k_\p = k_\m + \lambda k^{(1)}_\m + \mathcal{O}(\lambda^2). 
		\label{eq:ansatzFk}
\end{gather}
\eqref{eq:ansatzT} has to be fulfilled for every order of $\lambda$ separately. The zeroth order $T(0)=0$ is trivially fulfilled. The first order yields $\left.\mathrm{d}T/\mathrm{d}\lambda \right|_{\lambda=0}=0$, which results in
\begin{equation}
	k^{(1)}_\m \braket{\F_\m^\R| (k\Pop)^\prime |\F_\m}_\sV
	+
	k_\m \braket{\F_\m^\R| \Delta\Pop(k_\m)|\F_\m}_\sV
	+
	[\F_\m^\R | \F^{(1)}_\m]_{\spV}
	= 0
	\label{eq:firstOrder}.
\end{equation}
The prime denotes the derivative with respect to $k$, evaluated at $k_\m$. For the term $[\F_\m^\R | \F^{(1)}_\m]_{\spV}$, we make use of the fact that outside the inhomogeneity of the materials, the RSs can be expanded into a set of basis functions $\basis_\N$ that solve Maxwell's equations in homogeneous and isotropic space for outgoing boundary conditions~\cite{weiss2018scat}. The index $\N$ denotes a set of quantum numbers that labels the individual basis functions. As in Ref.~\cite{weiss2018scat}, by exploiting the $k$ dependence of the basis functions, we define an analytical continuation $\F_\m(k)$ and $\F_\p(k,\lambda)$ with $\F_\m(k_\m)=\F_\m$ and $\F_\p(k_\p,\lambda)=\F_\p$ in the exterior. Thus,
\begin{equation}
	[\F_\m^\R | \F^{(1)}_\m]_{\spV}
	=
	\left. \frac{\mathrm{d}}{\mathrm{d} \lambda} [\F_\m^\R | \F_\p(k, \lambda)]_{\spV}\right|_{\lambda=0}
	=
	\STl+k_\m^{(1)}[\F_\m^\R | \F^\prime_\m]_{\spV},
\label{eq:surfaceTermSplitted}
\end{equation}
where $\STl=\left. \partial/\partial \lambda [\F_\m^\R | \F_\p(k_m, \lambda)]_{\spV}\right|_{\lambda=0}$ and the last term arises due to the implicit $\lambda$ dependence of $k_\p$. Note that $\F^\prime_\m$ is the $k$ derivative of the analytic continuation $\F_\m(k)$ at $k_\m$. Inserting \eqref{eq:surfaceTermSplitted} into \eqref{eq:firstOrder}, we obtain the first-order expression for the change of the wavenumber as
\begin{equation}
k^{(1)}_\m 
=
-\frac
{ 
	k_\m\braket{\F_\m^\R| \Delta\Pop(k_\m)|\F_\m}_\sV + \STl
}
{
	\braket{\F_\m^\R| (k\Pop)^\prime |\F_\m}_\sV + [\F_\m^\R | \F^\prime_\m]_{\spV}
}
\label{eq:k1withEH}.
\end{equation}
This is exactly the same result as in Ref.~\cite{weiss2017oblique} with an additional contribution $\STl$ that allows for the description of a homogeneous perturbation in the homogeneous and isotropic exterior. 

Let us now evaluate the surface term $\STl$ for two highly relevant cases: (i) a system, in which the spatial inhomogeneity remains finite in all directions (e.g. a single nanoparticle), and (ii) a planar periodic system (e.g. a photonic crystal slab or an array of nanoantennas). It is straightforward to extend our approach to other geometries. For case~(i), we choose our integration surface $\partial V$ as a sphere that completely surrounds the inhomogeneity. For case~(ii), we split our integration surface $\partial V$ into two planes, one located above and one located underneath the inhomogeneity. As it is shown in Ref.~\cite{weiss2018scat}, it is possible in both cases to choose the basis functions $\basis_{\N}$ such that they fulfill the orthogonality relation $[\basis^\R_\N |  \basis_{\N^\prime}]_{\spV} = 0$ for all $\N$ and $\N^\prime$. Furthermore, the basis functions given in Ref.~\cite{weiss2018scat} can be factorized into the following form: 
\begin{equation}
\basis_\N(\es, \mus, k) = A_\N(\es, \mus, k) 	
\begin{pmatrix} \sqrt{\mus}\fE_\N(\ns k, \mathbf{r}) \\ i\sqrt{\es}\fH_\N(\ns k, \mathbf{r}) \\\end{pmatrix},
\end{equation}
where $A_\N(\es, \mus, k)$ is a normalization constant, $\es(k, \lambda) = \varepsilon(k)+\lambda\Delta\varepsilon(k)$, $\mus(k, \lambda) = \mu(k)+\lambda\Delta\mu(k)$,  $\ns(k, \lambda) = \sqrt{\es(k, \lambda) \mus(k, \lambda)}$, and $\fE_\N$ and $\fH_\N$ are vector functions that depend on the product of $\ns$ and  $k$. Using the basis functions, we can write the perturbed and unperturbed RSs as
$\F_\p(k, \lambda)=\sum_\N \alpha_\N(\lambda)\basis_\N[\es(k, \lambda), \mus(k, \lambda), k]$ and $\F_\m(k) =\sum_\N \alpha_\N(0)\basis_\N[\es(k, 0), \mus(k, 0), k]$, where $\alpha_\N(\lambda)$ are the perturbation-dependent expansion coefficients.
Inserting this into $\STl$, exploiting the orthogonality of $\basis_\N$, and making use of the relation $\oint_{\partial V } \mathrm{d}\mathbf{S}\cdot
\left( \mathbf{E}_\m \times  \mathbf{H}^\R_\m \right)=\oint_{\partial V } \mathrm{d}\mathbf{S}\cdot
\left( \mathbf{E}^\R_\m \times  \mathbf{H}_\m \right)$, we obtain
\begin{equation}
\STl =  \eta \frac{k_\m}{2} \left(\frac{\Delta\varepsilon}{\varepsilon}+\frac{\Delta\mu}{\mu}\right) \left[\F_\m^\R \left| \F^\prime_\m \right.\right]_{\spV}
+ 
\frac{i}{2} \eta \beta \oint_{\partial V } \mathrm{d}\mathbf{S}\cdot
\left( \mathbf{E}^\R_\m \times  \mathbf{H}_\m \right), 
\label{eq:ST}
\end{equation}
with the factor
\begin{equation}
\beta = \frac{(k \mu)^\prime}{\mu}\frac{\Delta\varepsilon}{\varepsilon}-\frac{(k\varepsilon)^\prime}{\varepsilon}\frac{\Delta\mu}{\mu},
\end{equation}
and the abbreviation $\eta = \sqrt{\varepsilon\mu}/(k\sqrt{\varepsilon\mu})^\prime$. Again, the prime denotes the $k$ derivative at $k_\m$. If not further specified, the material parameters $\varepsilon$, $\Delta \varepsilon$, $\mu$, and $\Delta \mu$ are meant to be taken at $k_\m$. Note that for non-dispersive materials, we trivially have $\varepsilon^\prime = \mu^\prime = 0$, $(k\mu)^\prime/\mu=(k\varepsilon)^\prime/\varepsilon=1$ and $\eta = 1$.

\eqref{eq:k1withEH} together with \eqref{eq:ST} are our final result and allow to calculate changes of the wavenumber as an integral expression over the unperturbed fields $\E_\m$ and $\Hv_\m$. While the occurring volume integrals are straightforward to evaluate, the surface integrals are a bit more sophisticated, due to the $k$ derivative $\F_\m^\prime$ appearing in $[\F_\m^\R | \F^\prime_\m]_{\spV}$. For the finite system~(i), we can get rid of the $k$ derivative by using the relation $\F^\prime_\m = 1/(\eta k_\m)(\rv \cdot \nabla) \F_\m$ (cf. Ref.~\cite{mul2018general}) . The evaluation of $[\F_\m^\R | \F^\prime_\m]_{\spV}$ for the planar periodic system~(ii) can be found in Ref.~\cite{weiss2018scat}.

We want to conclude our derivations with three additional remarks: First, our theory does not require any normalization of $\F_\m$. Instead, the analytical normalization condition derived in Ref.~\cite{mul2018general} is automatically contained in \eqref{eq:k1withEH} in the denominator. Second, \eqref{eq:k1withEH}, as well as its nominator and its denominator are independent of the size of the integration volume $V$, except that $V$ must enclose all spatial inhomogeneities. Third, in order to be as general as possible, we formulated our theory in terms of both the electric and the magnetic fields. For non-magnetic materials, it is possible to convert \eqref{eq:k1withEH} into an expression that contains only the electric field, by exploiting the relations provided in Ref.~\cite{mul2018general}.

 \begin{figure}[htb!]
	\centering
	\includegraphics[width=\linewidth]{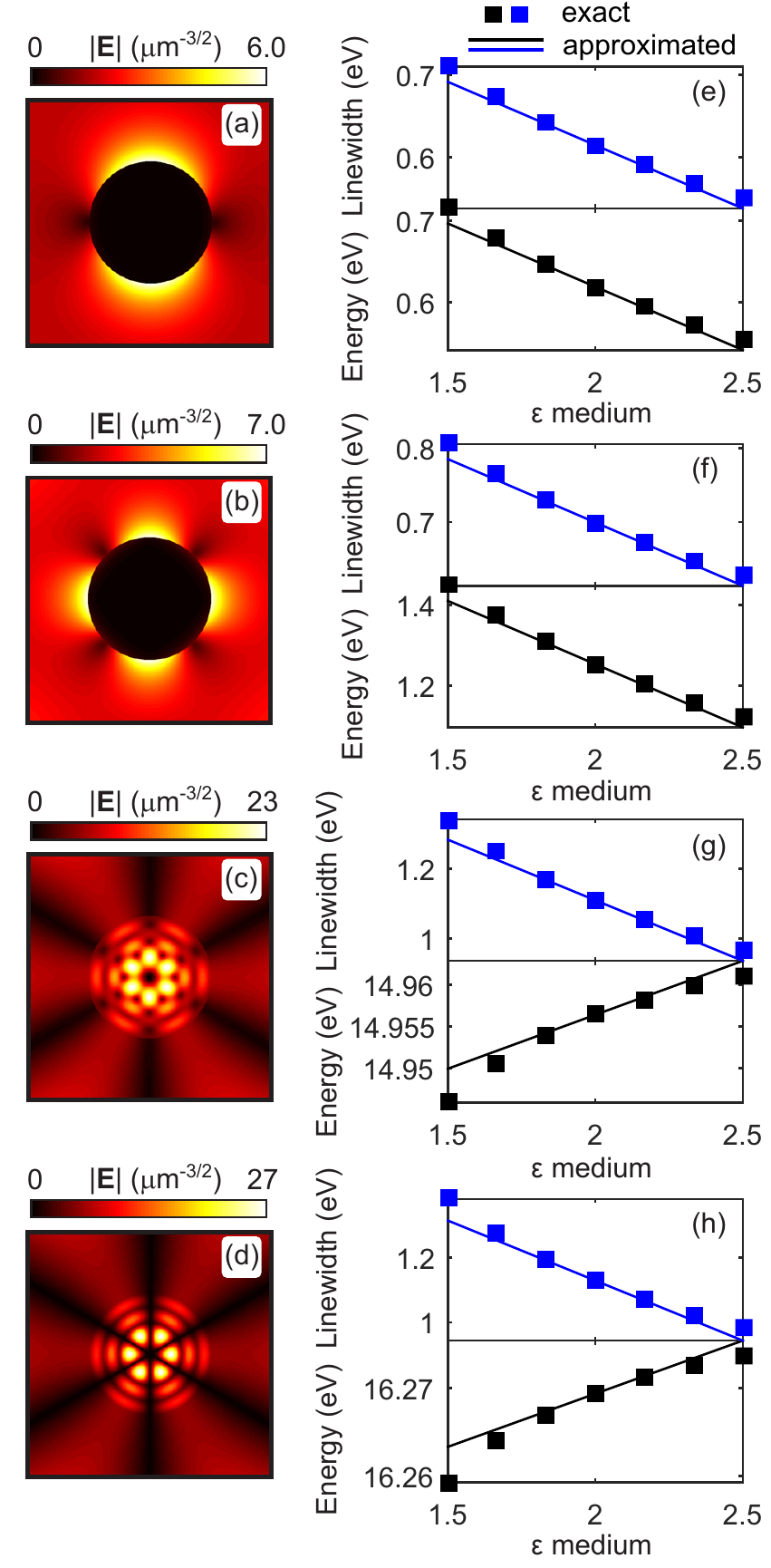}
	\caption{ Results for test system~(i). As illustrated in Fig.~\ref{fig:sphere_drawing}, we consider a gold nanosphere (diameter 400~nm) and vary the permittivity $\varepsilon$ of its surrounding medium. (a-d) Normalized electric field distribution of exemplary resonant states of the unperturbed system ($\varepsilon=2$). Panels (a) and (b) depict the fundamental plasmonic dipole and quadrupole mode, while (c) and (d) display higher-order transverse-magnetic and transverse-electric Mie resonances. (e-h) Resonance energy (black) and linewidth (blue) as a function of $\varepsilon$, with solid lines as the results of the first-order perturbation theory and squares derived by exact analytical calculations.}
	\label{fig:sphere_modes}
\end{figure}
Let us now test our theory at a simple example system: As depicted in Fig.~\ref{fig:sphere_drawing}, we consider a metallic nanosphere and vary the permittivity of its surrounding medium. The RSs of the sphere can be calculated analytically and are given in Refs.~\cite{DoostRSE3D2014, mul2016sand}. For our example, we take a gold sphere with a diameter of 400~nm, described by a Drude model ($\omega_{\text{p}}=13.8\times10^{15}~\text{s}^{-1}$ and $\gamma=1.075\times10^{14}~\text{s}^{-1}$). The unperturbed permittivity of the surrounding medium is chosen as $\varepsilon=2$.  

In Fig.~\ref{fig:sphere_modes} (a-d), we display the normalized electric field distribution of exemplary RSs of the unperturbed system. Panels (a) and (b) show the fundamental plasmonic dipole and quadrupole mode, which correspond to poles of the transverse-magnetic (TM) Mie coefficients for an angular momentum quantum number of $l=1$ and $l=2$, respectively, and occur at frequencies below the plasma frequency $\omega_{\text{p}}$, where the gold is metallic. For frequencies larger than $\omega_{\text{p}}$, the gold behaves as a dielectric, and whispering gallery modes inside the sphere are possible. Panels (c) and (d) show a transverse-magnetic (TM) and a transverse-electric (TE) higher-order whispering gallery mode, respectively, both with an angular momentum quantum number of $l=3$ and three radial antinodes inside the sphere.  Fig.~\ref{fig:sphere_modes}(e-h) depict the resonance energies (black) and linewidths (blue) of the four modes as a function of the permittivity $\varepsilon$ of the surrounding medium. The solid lines indicate the results of the perturbation theory, while the squares have been derived from exact analytical calculations~\cite{DoostRSE3D2014}. For not too big variations in $\varepsilon$, we have a good agreement between the linear perturbation theory and the exact calculations, while at the edge of the plotted $\varepsilon$ range, some deviations become visible.

\begin{figure}[htb]
	\centering
	\includegraphics[width=\linewidth]{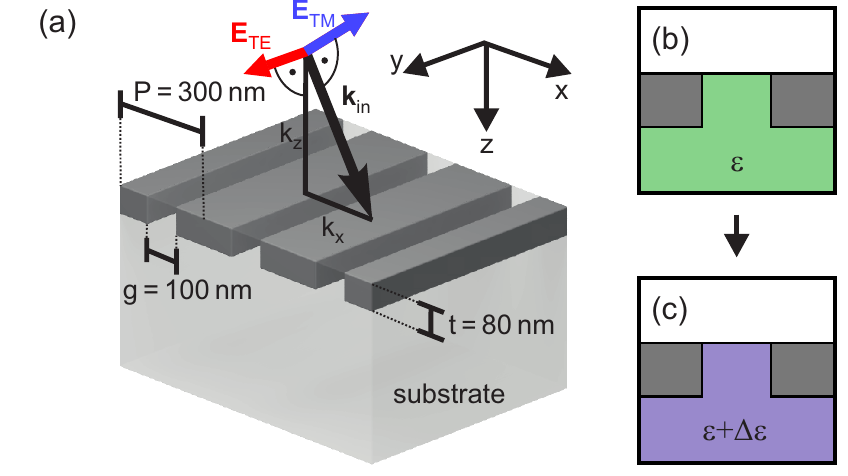}
	\caption{Schematic of test system~(ii). We consider a one-dimensional photonic crystal slab that was originally introduced in Ref.~\cite{Akimov2011} and further discussed in Ref.~\cite{weiss2017oblique}. (a) Structure geometry with parameters as specified in Refs.~\cite{weiss2017oblique,Akimov2011}. The structure consists of a periodic grating (dark gray) of a material with refractive index 2.5, embedded into a substrate with permittivity $\varepsilon$ (unperturbed case: $\varepsilon=2.25$), and air on top. (b,c) We introduce a perturbation by changing $\varepsilon$ to $\varepsilon+\Delta\varepsilon$.}
	\label{fig:grating_drawing}
\end{figure}
As a second example, we consider a one-dimensional photonic crystal slab, which was originally introduced in Ref.~\cite{Akimov2011} and was further discussed in Ref.~\cite{weiss2017oblique}. The geometry is depicted in Fig.~\ref{fig:grating_drawing}(a). The system is periodic in the $x$ direction, translationally symmetric in the $y$ direction, and remains finite within the $z$ direction. It consists of a 80~nm thick periodically modulated layer (period ${P=300~\text{nm}}$) with a 200~nm wide region of ZnO (dark gray, ${n = 2.5}$) per unit cell, embedded into a quartz substrate (light gray) with a permittivity value $\varepsilon$, where ${\varepsilon=2.25}$ in the unperturbed case, and an air cover layer. As indicated in Fig.~\ref{fig:grating_drawing}(b,c), we change the permittivity of the quartz from $\varepsilon$ to $\varepsilon+\Delta\varepsilon$.
\begin{figure}[htb]
	\centering
	\includegraphics[width=\linewidth]{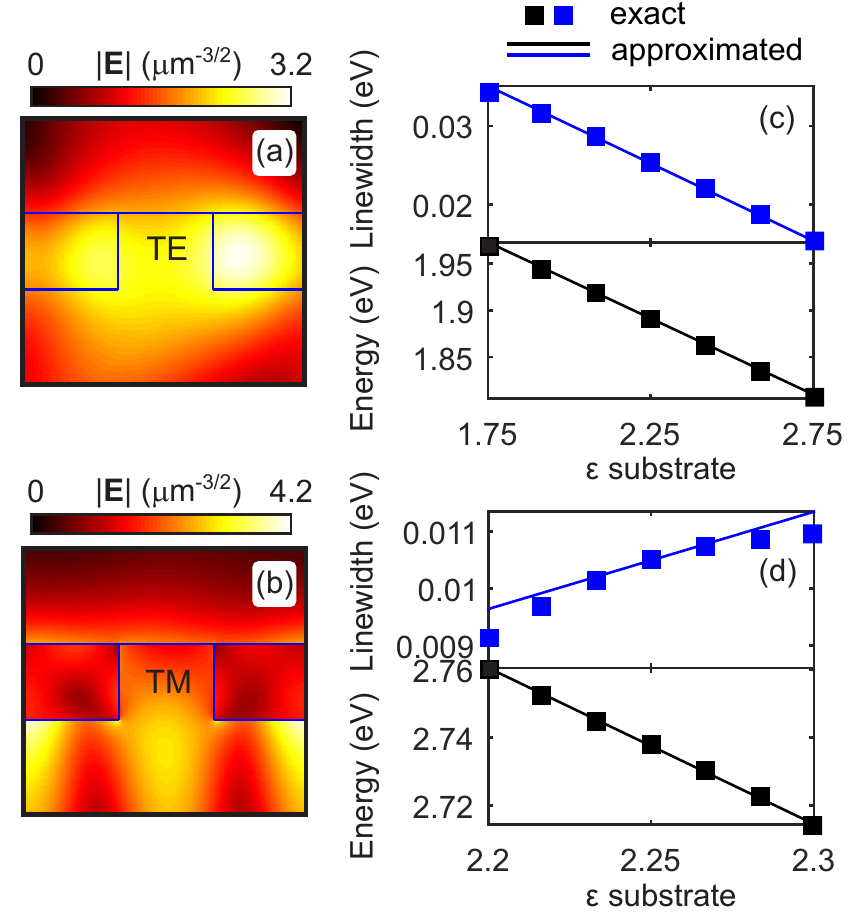}
	\caption{Results for test system~(ii). (a,b) Normalized electric field distributions of exemplary resonant states within the unperturbed system ($\varepsilon=2.25$) that are also considered in Ref.~\cite{weiss2017oblique}: (a) Transverse-electric (TE) mode at $k_x = \pi/(2P) = 5.236~\text{\textmu m}^{-1}$, (b) Transverse-magnetic (TM) mode at $k_x = 0.2~\text{\textmu m}^{-1}$. (c,d) Corresponding resonance energy (black) and linewidth (blue) as a function of the permittivity $\varepsilon$ of the substrate. Solid lines represent the results of the first-order perturbation theory, while the squares have been derived by numerically exact calculations. Note that (c) and (d) are plotted for different ranges of $\varepsilon$.}
	\label{fig:grating_modes}
\end{figure}
The RSs of the system correspond to quasiguided TE and TM waveguide modes~\cite{Tikhodeev2002quasi, weiss2017oblique}. Due to the periodicity, the RSs can be written as Bloch waves, which are characterized by their in-plane momentum $k_x$. As in Ref.~\cite{weiss2017oblique}, the field distribution of the RSs in the unperturbed system, as well as the exact resonance frequencies in the perturbed case, have been calculated using the Fourier modal method~\cite{weiss2016perturbation, weiss2009matchedCoordinates, weiss2011plasmonicResonancesFMM, bykov2013scatteringMatrix}. Exploiting the periodicity and the translational symmetry~\cite{weiss2017oblique}, the calculation domain, as well as the integration volume $V$ appearing in \eqref{eq:k1withEH}, can be reduced to a two-dimensional rectangle within the $xz$ plane that spans over one unit cell in the $x$ direction and covers the inhomogeneity in the $z$ direction.

Fig.~\ref{fig:grating_drawing}(a,b) show the normalized electric field distribution of exemplary RSs in the unperturbed system. The example uses exactly the same modes as discussed in Ref.~\cite{weiss2017oblique}, which are a TE resonance at $k_x = \pi/(2P) = 5.236~\text{\textmu m}^{-1}$ (a), and a TM resonance at $k_x = 0.2~\text{\textmu m}^{-1}$ (b). Panels (c) and (d) depict the corresponding resonance energy (black) and linewidth (blue) as a function of $\varepsilon$. The solid lines represent the result of the first-order perturbation theory, while the squares have been derived from exact numerical calculations. For both modes, perturbation theory and exact results exhibit a good agreement, as long as the change in $\varepsilon$ is not too big. Note that for the TE mode, the linear perturbation theory works over a much larger range of $\varepsilon$ than for the TM mode. The reason is that the TM resonance depicted here is coincidentally very close to a Rayleigh anomaly~\cite{weiss2017oblique} that strongly effects the far field coupling, which in turn significantly depends on the substrate index that is changed here as the perturbation parameter.

In conclusion, we have generalized the single-mode approximation of the resonant state expansion to perturbations in the exterior of open optical resonators. The key is to include an additional surface term that describes the changes in the surrounding. Explicit expressions as well as exemplary validations are given for two practically important cases: Single nanoparticles and periodic structures. We believe that our theory extends the capabilities of the resonant state expansion as an efficient toolbox for modeling and designing nanophotonic systems.

\paragraph*{Funding.}

Deutsche Forschungsgemeinschaft (DFG SPP 1839); VW Foundation; Ministerium für Wissenschaft, Forschung und Kunst Baden-Württemberg (MWK).

\bibliography{paper}

\begin{thebibliography}{10}
\newcommand{\enquote}[1]{``#1''}

\bibitem{wilson2002}
W.~D. Wilson, {\protect\JournalTitle{Science}} \textbf{295}, 2103 (2002).

\bibitem{arnold2003}
S.~Arnold, M.~Khoshsima, I.~Teraoka, S.~Holler, and F.~Vollmer,
  {\protect\JournalTitle{Opt. Lett.}} \textbf{28}, 272 (2003).

\bibitem{unger2009}
A.~Unger and M.~Kreiter, {\protect\JournalTitle{The Journal of Physical
  Chemistry C}} \textbf{113}, 12243 (2009).

\bibitem{liu2010}
N.~Liu, T.~Weiss, M.~Mesch, L.~Langguth, U.~Eigenthaler, M.~Hirscher,
  C.~Sönnichsen, and H.~Giessen, {\protect\JournalTitle{Nano Letters}}
  \textbf{10}, 1103 (2010). PMID: 20017551.

\bibitem{cetin2014}
A.~E. Cetin, A.~F. Coskun, B.~C. Galarreta, M.~Huang, D.~Herman, A.~Ozcan, and
  H.~Altug, {\protect\JournalTitle{Light: Science \&Amp; Applications}}
  \textbf{3}, e122 EP  (2014). Original Article.

\bibitem{gallinet2013}
B.~Gallinet, T.~Siegfried, H.~Sigg, P.~Nordlander, and O.~J.~F. Martin,
  {\protect\JournalTitle{Nano Letters}} \textbf{13}, 497 (2013).

\bibitem{meschFano2018}
M.~Mesch, T.~Weiss, M.~Sch{\"a}ferling, M.~Hentschel, R.~S. Hegde, and
  H.~Giessen, {\protect\JournalTitle{ACS Sensors}} \textbf{3}, 960 (2018).

\bibitem{lalanne2015simple}
J.~Yang, H.~Giessen, and P.~Lalanne, {\protect\JournalTitle{Nano Letters}}
  \textbf{15}, 3439 (2015). PMID: 25844813.

\bibitem{hughes2018reg}
M.~Kamandar~Dezfouli and S.~Hughes, {\protect\JournalTitle{Phys. Rev. B}}
  \textbf{97}, 115302 (2018).

\bibitem{lalanne2018review}
P.~Lalanne, W.~Yan, K.~Vynck, C.~Sauvan, and J.-P. Hugonin,
  {\protect\JournalTitle{Laser \& Photonics Reviews}} \textbf{12}, 1700113
  (2018).

\bibitem{colom2018}
R.~Colom, R.~McPhedran, B.~Stout, and N.~Bonod, {\protect\JournalTitle{Phys.
  Rev. B}} \textbf{98}, 085418 (2018).

\bibitem{DoostRSE3D2014}
M.~B. Doost, W.~Langbein, and E.~A. Muljarov, {\protect\JournalTitle{Phys. Rev.
  A}} \textbf{90}, 013834 (2014).

\bibitem{zhang2015univ}
W.~Zhang and O.~J.~F. Martin, {\protect\JournalTitle{ACS Photonics}}
  \textbf{2}, 144 (2015).

\bibitem{mul2016sand}
E.~A. Muljarov and W.~Langbein, {\protect\JournalTitle{Phys. Rev. B}}
  \textbf{93}, 075417 (2016).

\bibitem{weiss2016perturbation}
T.~Weiss, M.~Mesch, M.~Sch\"aferling, H.~Giessen, W.~Langbein, and E.~A.
  Muljarov, {\protect\JournalTitle{Phys. Rev. Lett.}} \textbf{116}, 237401
  (2016).

\bibitem{weiss2017oblique}
T.~Weiss, M.~Sch\"aferling, H.~Giessen, N.~A. Gippius, S.~G. Tikhodeev,
  W.~Langbein, and E.~A. Muljarov, {\protect\JournalTitle{Phys. Rev. B}}
  \textbf{96}, 045129 (2017).

\bibitem{upendar2018}
S.~Upendar, I.~Allayarov, M.~A. Schmidt, and T.~Weiss,
  {\protect\JournalTitle{Opt. Express}} \textbf{26}, 22536 (2018).

\bibitem{koenderink2010purcell}
A.~F. Koenderink, {\protect\JournalTitle{Opt. Lett.}} \textbf{35}, 4208 (2010).

\bibitem{mul2018general}
E.~A. Muljarov and T.~Weiss, {\protect\JournalTitle{Opt. Lett.}} \textbf{43},
  1978 (2018).

\bibitem{weiss2018scat}
T.~Weiss and E.~A. Muljarov, {\protect\JournalTitle{Phys. Rev. B}} \textbf{98},
  085433 (2018).

\bibitem{Akimov2011}
A.~B. Akimov, N.~A. Gippius, and S.~G. Tikhodeev, {\protect\JournalTitle{JETP
  Letters}} \textbf{93}, 427 (2011).

\bibitem{Tikhodeev2002quasi}
S.~G. Tikhodeev, A.~L. Yablonskii, E.~A. Muljarov, N.~A. Gippius, and
  T.~Ishihara, {\protect\JournalTitle{Phys. Rev. B}} \textbf{66}, 045102
  (2002).

\bibitem{weiss2009matchedCoordinates}
T.~Weiss, G.~Granet, N.~A. Gippius, S.~G. Tikhodeev, and H.~Giessen,
  {\protect\JournalTitle{Opt. Express}} \textbf{17}, 8051 (2009).

\bibitem{weiss2011plasmonicResonancesFMM}
T.~Weiss, N.~A. Gippius, S.~G. Tikhodeev, G.~Granet, and H.~Giessen,
  {\protect\JournalTitle{J. Opt. Soc. Am. A}} \textbf{28}, 238 (2011).

\bibitem{bykov2013scatteringMatrix}
D.~A. Bykov and L.~L. Doskolovich, {\protect\JournalTitle{Journal of Lightwave
  Technology}} \textbf{31}, 793 (2013).

\end{thebibliography}

 


\end{document}